\documentclass[sigconf]{acmart} 

\newcommand{\my}[1]{{\color{black}{#1}}}

\usepackage{diagbox}
\usepackage{graphicx}
\usepackage{amsmath}
\usepackage{amsfonts}
\usepackage{algpseudocode}
\usepackage{algorithm}
\usepackage{subfigure}

\newtheorem{example}{Example}

\begin{document}
\title{Towards Safe Machine Learning for CPS}
\subtitle{Infer Uncertainty from Training Data}

\author{Xiaozhe Gu}
\affiliation{\institution{Energy Research Institute @ Nanyang Technological University}}
\email{XZGU@ntu.edu.sg}

\author{Arvind Easwaran}
\affiliation{\institution{School of Computer Science and Entineering, Nanyang Technological University}}
\email{arvinde@ntu.edu.sg}


\begin{abstract}
Machine learning (\emph{ML})  techniques are increasingly applied to decision-making and control problems in Cyber-Physical Systems   among which many are safety-critical, e.g., chemical  plants, robotics, autonomous vehicles.  Despite the significant benefits brought by  ML techniques, they also raise additional safety issues because  1) most expressive and  powerful ML models are not transparent and behave as a black box and 2) the training data which  plays a crucial role in ML safety  is usually incomplete. An important  technique   to achieve safety for ML models is ``Safe Fail'', i.e., a model selects a reject option and applies the backup solution, a traditional controller or a human operator for example, when it has low confidence in a prediction.  

Data-driven models produced by ML algorithms learn from training data,  and hence they are only as good as the examples they have learnt.  \my{As pointed in~\cite{marcus2018deep}, ML models work well in the ``\emph{training space}'' (i.e., feature space  with sufficient training data), but they could not extrapolate beyond the training space.}   As observed in many previous studies, a feature space that lacks training data generally has a much higher error rate than the one that contains sufficient training samples~\cite{weiss2004mining}. Therefore, it is essential  to identify the training space  and  avoid extrapolating beyond the training space.  In this paper, we propose an efficient Feature Space Partitioning Tree (\emph{FSPT}) to address this problem. Using experiments, we also show that,  \my{a strong relationship exists between model performance and FSPT score.}
\end{abstract}

%
%



\copyrightyear{2019} 
\acmYear{2019} 
\setcopyright{acmlicensed}
\acmConference[ICCPS '19]{10th ACM/IEEE International Conference on Cyber-Physical Systems (with CPS-IoT Week 2019)}{April 16--18, 2019}{Montreal, QC, Canada}
\acmBooktitle{10th ACM/IEEE International Conference on Cyber-Physical Systems (with CPS-IoT Week 2019) (ICCPS '19), April 16--18, 2019, Montreal, QC, Canada}
\acmPrice{15.00}
\acmDOI{10.1145/3302509.3311038}
\acmISBN{978-1-4503-6285-6/19/04}

\maketitle

\section{Introduction}
Cyber-physical systems (CPS) are the new generation of engineered systems that continually interact with  the physical world and human operators. Sensors, computational and physical processes  are all tightly coupled together in CPS. Many CPS have already been deployed in safety-critical domains such as aerospace, transportation, and healthcare.

On the other hand, machine learning (ML) techniques have achieved impressive results in recent years. They can reduce development cost as well as  provide   practical solutions to complex tasks  which cannot be solved by traditional methods. Not surprisingly, ML techniques have been applied to many decision-making and control problems in CPS  such as energy control~\cite{jain2018learning}, surgical robots~\cite{lin2005automatic}, self-driving~\cite{bojarski2016end}, and so forth.  The safety-critical nature of CPS involving ML raises the need to improve system safety and reliability.  Unfortunately,    ML has many undesired  characteristics that can impede this achievement of safety and reliability. 
\begin{itemize}
  \item   ML models with strong expressive power, e.g., deep neural networks (DNN), are typically considered non-transparent. Non-transparency is an obstacle to safety assurance because if the  model behaves as a black box and cannot be understood by an assessor,  it is  difficult to develop confidence that the model is operating as intended. 
  \item   The standard empirical risk minimization approach used to train ML models reduces the empirical loss of a subset of possible inputs (i.e., training samples) that could be encountered operationally.   An implicit assumption made here is that training samples are drawn based on the actual underlying probability distribution.  As a result,  the representativeness of training samples is a necessary condition to produce reliable ML models. However,  this may not always be the case, and training samples could be absent from most parts of the feature space.

\end{itemize}
We can apply various techniques to improve the safety/reliability of ML models~\cite{safeml}. To increase  transparency, we can insist on models that can be interpreted by people such as  ensembles of low-dimensional interpretable sub-models~\cite{nusser2008interpretable} or use specific explainers~\cite{lime} to interpret the predictions made by ML models.  We can also  exclude features~\cite{Udeshi_2018} that are not causally related to the outcome.  A practical technique for ML to avoid unsafe predictions is ``Safe Fail''. If a model is not likely to produce a correct output, a reject option is selected,  and  the backup solution, a traditional  non-ML approach or  a human operator, for example,  is applied, thereby causing the system to fail safely.  \my{Such a ``Safe Fail'' technique is also not new in ML based system~\cite{bartlett2008,herbei2006classification}.  These works~\cite{bartlett2008,herbei2006classification} focus on minimizing the empirical loss of the training set and hence implicitly assume the training set to be representative.  For example, a support vector machine (SVM) like classifier  with a reject option~\cite{bartlett2008}  could be used for this purpose.}  As shown in  Equation~\ref{eq:reject}, $\phi(\mathbf x)$ is the predictive output of the SVM classifier, which is interpreted as its confidence in a prediction, and $t$ is the threshold for the  reject option. The classifier is supposed to be most uncertain when $\phi(\mathbf x)$ approximates to $0$.
\begin{align}
\label{eq:reject}
\hat y(\mathbf x)=\begin{cases}-1~\mbox{ if  }&\phi(\mathbf x)\leq -t\\\mbox{reject option, if }&\phi(\mathbf x)\in(-t,t)\\1~\mbox{     ~~if  }&\phi(\mathbf x)\geq t\end{cases}
\end{align}
\emph{As we can observe from  Equation~\ref{eq:reject}, to determine a reject option, we need the  prediction ``confidence''  $\phi(\mathbf x)$, which however, can be misleading.}

\subsection{Motivation}
\label{sec:moti}
\begin{figure}[t]
\centering
\includegraphics[width=0.5\textwidth]{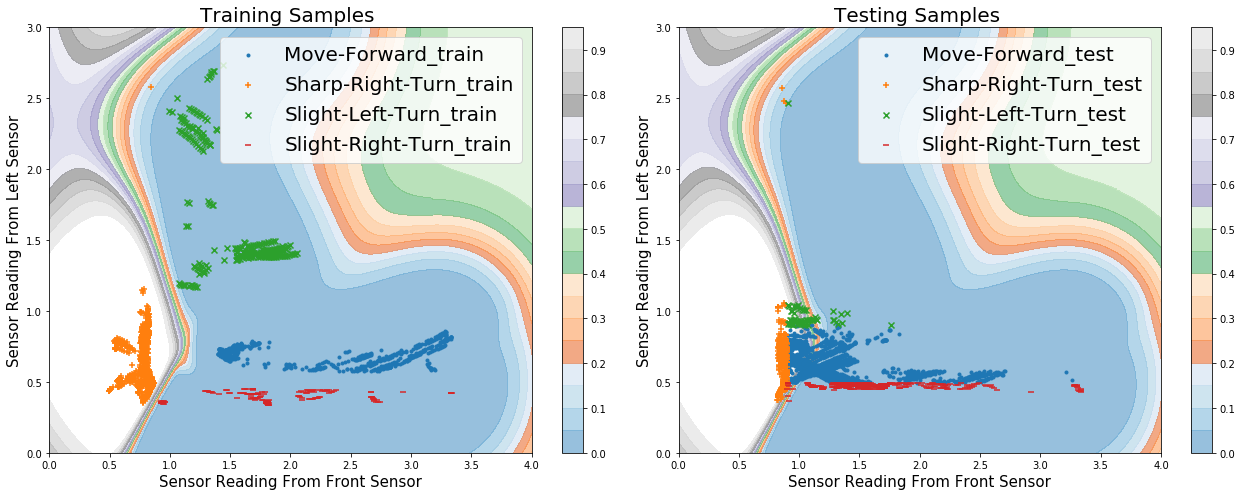}
\caption{Wall-following navigation task with mobile robot SCITOS-G5 based on sensor readings from the front and left sensor~\cite{Dua:2017}.}
\label{fig:toy1}
\end{figure}

\begin{figure}[t]
\centering
\includegraphics[width=0.33\textwidth]{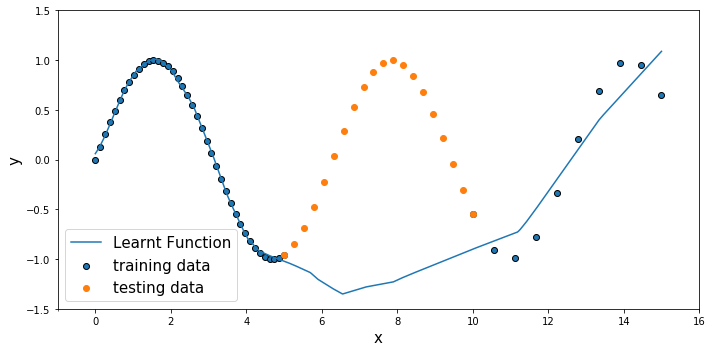}
\caption{A toy regression example }
\label{fig:toyx}
\end{figure}

ML  models learn from a subset of possible scenarios that could be encountered operationally.   Thus, they can only be as good as the training examples they have learnt.   \my{As pointed in~\cite{marcus2018deep}, ML models work well in the ``training space''  with a cloud of training points, but they could not extrapolate beyond the training space.   In other words,  the training data determines the training space and hence the upper bound of ML model's performance.  A previous study~\cite{weiss1995learning} has  also demonstrated that  a feature space that lacks training data generally has a much higher error rate.   Unfortunately, the training data is usually incomplete in practice and covers a very small part of the entire feature space. In fact, there is no guarantee that the training data is even representative~\cite{ISO16}. Here we use two simple examples to illustrate this problem.}
\begin{example}
   Figure~\ref{fig:toy1} shows the decision boundaries of an SVM classifier to predict  whether a mobile robot is turning right sharply.  The value in the contour map represents the ``predictive probability''  that the input instance belongs to the class ``Sharp-Right-Turn''.  In this example, the training samples are not  representative of testing samples, \my{and only cover a very small portion of the feature space.}  However,  the classifier still has very high confidence beyond its training space  even though  there exists no training samples. As a result, the accuracy of testing samples decreases to $66\%$ while the accuracy of training samples is almost $100\%$.
    \label{example:toy1}
\end{example}
\begin{example}
\my{
Figure~\ref{fig:toyx} shows a toy regression problem, where 40 training samples drawn from a sine function have feature x in $[0,5]$, and 10 training samples have feature x in  $(10,15]$.  However, the testing samples have feature x in  $(5,10]$.  We use a neural network (NN)  regressor to fit the data, and as shown, NN does a better job in fitting the sine function in $[0,5]$ than in $(10,15]$.  Meanwhile,  it does a terrible job in  extrapolating outside of the training space, i.e., $(5,10]$.
}
\label{example:toy2}
\end{example}

\subsection{Contribution}
\my{
From the above examples we can observe that ML models work well only in the ``training space'' (i.e., feature space with sufficient training samples.) Meanwhile,  in general,  ML models have greater potential to achieve better performance in feature space with more training samples.  Therefore,  it is essential for safety-critical ML based systems to ensure that their underlying ML modules   only  work in the training space.  As a result, we aim to design a novel technique to do the following job:

}
\begin{enumerate}
  \item split the feature space into multiple partitions, 
  \item identify those in which training samples are insufficient, and
  \item \my{reject input instances from these data-lacking feature space partitions.}  
\end{enumerate}
We first outline  the \textbf{desired characteristics}  for such a technique:
\begin{enumerate}
    \item   A score function is required to evaluate the resulting feature space partitions.
    \item   The  boundaries of feature space partitions are preferred to be interpretable and  understandable, so that we can  know, in which partition, ML models may have poor performance. With such information, we can collect more training samples from these regions  (if possible).
    \item Since we use this technique as a complement to  ML models,  the output must be generated efficiently, and the additional overhead should be as small as possible.

\end{enumerate}
In this paper, we propose a  Feature Space Partitioning Tree (FSPT)  with the characteristics  mentioned above, which comprises  

\begin{figure}[t]
\centering
\includegraphics[width=0.38\textwidth]{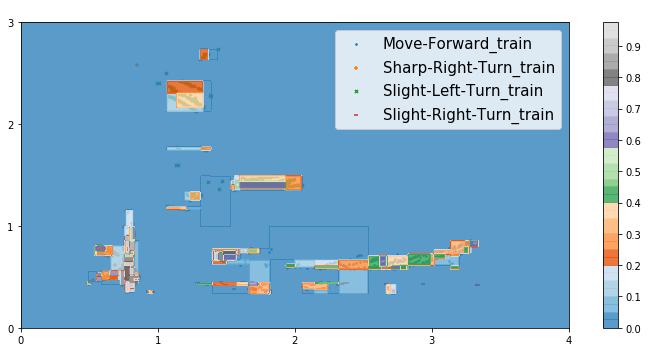}
\caption{ The resulting feature space partitions with  scores from FSPT for classification problem in Example~\ref{example:toy1}}
\label{fig:toy2}
\end{figure}

\begin{enumerate}
  \item a tree-based classifier for splitting feature space (Section~\ref{sec:tree}) with specific stopping and splitting criterion~(Section~\ref{sec:stop} to \ref{sec:newsplit}), and 
  \item a score function $S(\mathcal R)$ to evaluate the resulting feature space partitions (Section~\ref{sec:score}).
\end{enumerate}
As a toy example, Figure~\ref{fig:toy2} shows  the resulting feature space partitions for  Example~\ref{example:toy1}. The color of each hyper-rectangle represents the scores from FSPT. As we can see, \my{FSPT gives very low score to most partitions of the feature space because the training data only covers a small portion of the feature space\footnote{\my{Note that,  it does not mean that we have to  reject most input instances. In fact, if the training data is representative, we will encounter few input instances from these low score feature space partitions.}}.}

\noindent \textbf{Organization}: In Section~\ref{sec:related_work} we present related work, and in Section~\ref{sec:tree}, we propose the technique  for splitting the feature space into multiple hyper-rectangles based on Classification And Regression Tree (CART)~\cite{Breiman:2253780}.  We propose customization techniques (e.g., a new splitting criterion that takes feature importance into consideration and a score function for the resulting feature space partitions) in Section~\ref{sec:cust}.   In Section~\ref{sec:rej}, we introduce  how to apply FSPT in  ML with a reject option.  Finally, the experimental results in Section~\ref{sec:evaluation} also meet our expectations that, on average, ML models have a  higher loss/error rate in feature space partitions with lower FSPT scores.  In Table~\ref{tab:notation}, we list notations that will be used in the remainder of this paper.


\begin{table}[h]
\caption{{Notations}}
\label{tab:notation}
\footnotesize
\center
\begin{tabular}{|l|l|l|l|l|l|l|}
  \hline
$\mathbf Z=(\mathbf X,\mathbf y)$ &   data set\\
 \hline
 $\mathbf X$ &   $N\times d$ feature matrix with $N$ samples and $d$ dimensions\\
  \hline
      $y$&  label of an input instance\\
    \hline
      $\hat y$&  prediction  of an input instance\\
  \hline
     $\mathbf y$ &   vector of labels \\
  \hline
    $\mathbf x$&  feature vector of an input instance\\
  \hline
  $\mathbf I$ & a particular feature index\\
  \hline
    $f_{\mathbf I}$ & importance of feature   $\mathbf I$\\
  \hline
      $\mathbf x_{\mathbf I}^k$&  the value of the kth  sample in $\mathbf X$  on feature   $\mathbf I$\\
  \hline
  $\mathcal R$&  a feature space partition\\
  \hline
   $\overline{\mathbf I}(\mathcal R)$&  upper bound  value of  feature $\mathbf I$  of $\mathcal R$\\
  \hline
     $\underline{\mathbf I}(\mathcal R)$& lower bound  value of  feature $\mathbf I$  of $\mathcal R$ \\
       \hline
$\Delta \mathbf I(\mathcal R)$ & $\overline{\mathbf I}(\mathcal R)-\underline{\mathbf I}(\mathcal R)$\\
  \hline
       $|\mathcal R^+|$&  weighted number of training samples in $\mathcal R$ \\
  \hline
         $|\mathcal R^-|$&  weighted number of E-points in $\mathcal R$ \\
\hline
$G(\mathcal R)$& Gini index of $\mathcal R$\\
\hline
$\hat G(\mathcal R,\mathbf I,s)$& weighted Gini index with split feature $\mathbf I$ and value $s$\\
\hline
$\Delta G(\mathcal R)$& gain in Gini index\\
\hline
$S(\mathcal R)$ &  score of a resulting feature space partition $\mathcal R$ \\
\hline
$\phi_F(\mathbf x)$ &  output of FSPT for input instance $\mathbf x$ \\
\hline
$\phi_M(\mathbf x)$ &  output of the ML model for input instance $\mathbf x$ \\
\hline
\end{tabular}
\end{table}

\section{Related Work}
\label{sec:related_work}

\my{
In order to determine whether an ML model  should select a reject option,  we must obtain the ``confidence'' in its predictive output. } For classification problems,  the output of an ML model is usually interpreted as its confidence in that prediction. For example,  the output obtained at the end of the softmax layers of standard deep learning are often interpreted as the predictive probabilities.   \my{For ensemble methods\footnote{Ensemble methods are learning algorithms that construct a set of  classifiers and then classify new data points by taking a weighted vote of their predictions.}, the weighted votes of the  underlying  classifiers~\cite{varshney2013practical} will be used as the predictive probabilities.   As a result, existing works on classification with a reject option~\cite{bartlett2008,herbei2006classification} usually determine a reject option based on the predictive output. }

An implicit assumption made here is that these classifiers are most uncertain near the decision boundaries of different classes and the distance from the decision boundary is inversely related to the confidence that an input instance belongs to a particular class. This assumption is reasonable  in some sense  because the decision boundaries learnt by these models are usually located where many training samples with different labels overlap.   However, if a feature space $\mathcal X$ contains few or no training samples at all, then the decision boundaries of ML models may wholly be based on an inductive bias, thereby having much epistemic uncertainty~\cite{Attenberg:2015}. In other words, it is possible that an input instance coming from a feature space partition without any training samples would be classified falsely with a very high ``predictive probability'' by the ML model~\cite{gal2016dropout}. 

Prediction confidence can also be obtained by Bayesian methods.  \my{Unlike classical learning algorithm,  Bayesian algorithms do not attempt to identify ``best-fit'' models of the data. Instead, they compute a posterior distribution over models $P(\theta|\mathbf X,\mathbf y)$. 
For example, Gaussian process (GP)~\cite{seeger2004gaussian} assumes  that $p(f(\mathbf x^1), \ldots, f(\mathbf x^N ))$ is jointly Gaussian $\mathcal N(\mathbf \mu, \Sigma)$.  Given unobserved instance $\mathbf x^*$, the output  $\hat f(\mathbf x^*)$ of GP is then also conditional Gaussian $p(\hat f(\mathbf x^*)|\mathbf x^*,\mathbf X,\mathbf y)=\mathcal N(\mu_*,\sigma_*)$}.  The standard deviations $\sigma_*$ can then be interpreted as the prediction uncertainty. GP is  computationally intensive and has complexity $O(N^3)$, where N is the number of training samples.  Bayesian methods can also be applied to neural networks (NNs). Infinitely wide single hidden layer NNs with distributions placed over their weights converge to Gaussian processes~\cite{neal2012bayesian}. Variational inference~\cite{paisley2012variational,kingma2013auto} can be used to obtain  approximations for finite Bayesian neural networks. The  dropout  techniques in NNs can also be interpreted as a Bayesian approximation of  Gaussian process~\cite{gal2016dropout}. Despite the nice properties of Bayesian inference,  \emph{there are some   controversial aspects}: 
\my{
\begin{enumerate}
    \item The  prior plays a key role  in computing the marginal likelihood because  we are averaging the likelihood over all possible parameter settings $\theta$, as weighted by the prior~\cite{robert2014machine}. If the prior is not carefully chosen,  we may generate misleading results.
    \item It often comes with a high computational cost, especially in models with a large number of parameters.
\end{enumerate}
}

The conformal prediction framework~\cite{Vovk:2005,vovk2009line} uses past experience to determine precise levels  of confidence in new predictions. Given a certain error probability  requirement $\epsilon$, it forms a prediction interval $[\overline{f(\mathbf x)},\underline{{f( \mathbf x)}}]$ for regression or a  prediction label set $\{\mbox{Label 1},\mbox{Label 2},\ldots \}$ for classification  so that the interval/set contains the actual prediction with a probability greater than $1-\epsilon$. However, \emph{its theoretical correctness depends on the assumption that all the data are independent and identically distributed (later, a weaker assumption of ``exchangeability'' replaces this  assumption)}.  Besides, for regression problems, it tends to  produce prediction bands  whose width are roughly constant over the whole feature space~\cite{lei2018distribution}.


\section{Basic Idea of Partitioning the  Feature Space}
\label{sec:tree}
\begin{figure}[t]
\centering
\begin{minipage}[t]{0.22\textwidth}
  \centering
  \includegraphics[width=\textwidth]{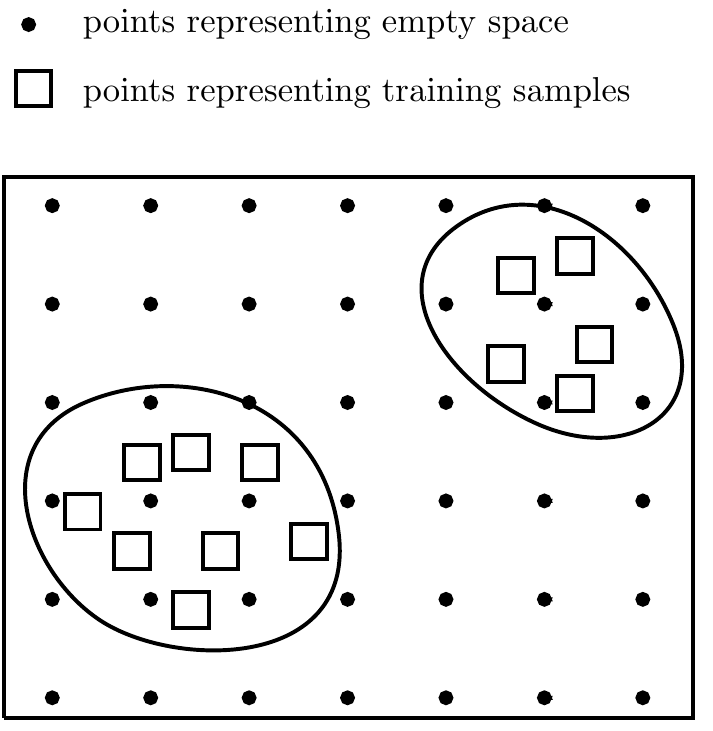}
\caption{Decision boundaries of a classifier}
\label{fig:basic1}
\end{minipage}
\begin{minipage}[t]{0.22\textwidth}
  \centering
  \includegraphics[width=\textwidth]{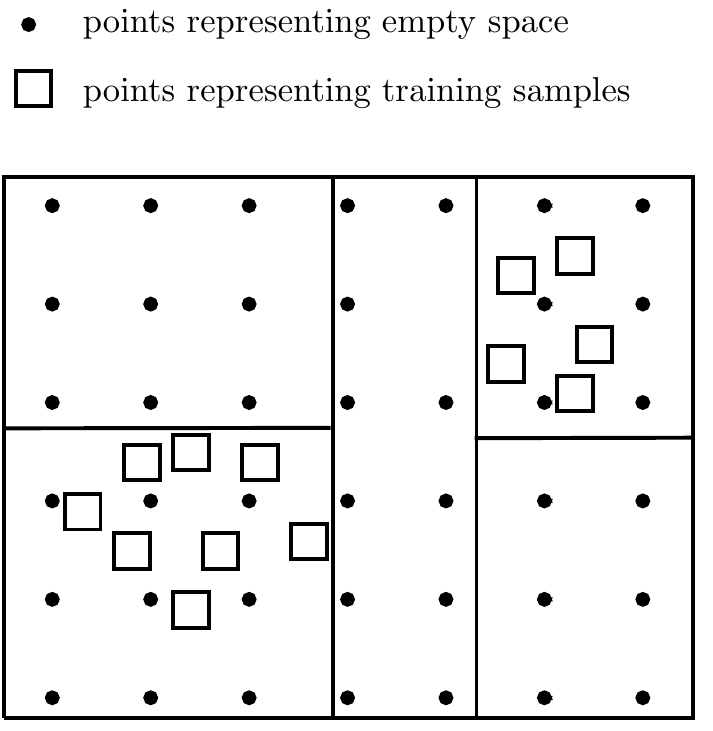}
\caption{Decision boundaries of a  tree-based classifier}
\label{fig:result_tree}
\end{minipage}
\end{figure}

\my{Our objective is to distinguish the feature space partitions with a high density of training samples from those with  a low  density of training samples.   Let's assume that there is another category of data points representing the empty feature space (\emph{E-points} for short)   that are uniformly distributed among the entire feature space. As shown in Figure~\ref{fig:basic1}, we can use a classifier to  distinguish the training data points from E-points. Then, the output of the classifier can be used to indicate whether an input instance is from a  feature space partition with sufficient training data.

}



However, the number of E-points we need to sample will increase exponentially with the number of features. There are two possible solutions to address this  issue:
\begin{enumerate}
\item apply dimension reduction techniques such as ensembles of low-dimensional sub-models~\cite{nusser2008interpretable}  or 
\item use tree/rule-based classifiers because their following properties are suitable for our task. 
\begin{enumerate}
    \item As shown in  Figure~\ref{fig:result_tree}, the feature space partitions constructed by a tree-based classifier are hyper-rectangles. As a result,  we can get useful information about each partition, e.g., the side length of each feature, the volume,  and the number of training samples within it,   very easily.
    \item Suppose the initial feature space $\mathcal R$ is split into  $\mathcal R_1$ and $\mathcal R_2$ by  feature $\mathbf I$ and value $s$, such that  $\forall \mathbf x \in \mathcal R_1:\mathbf x_{\mathbf I}\leq s$ and   $\forall \mathbf x \in \mathcal R_2:\mathbf x_{\mathbf I}> s$. The number of E-points belonging  to $\mathcal R_1$  and $\mathcal R_2$  will be in proportion to  the side length  of feature $\mathbf I$, i.e.,  $\Delta \mathbf I(\mathcal R_1)=s-\underline{\mathbf I}(\mathcal R)$ and  $\Delta \mathbf I(\mathcal R_2)=\overline{\mathbf I}(\mathcal R)-s$ because the  E-points are assumed to be uniformly distributed.
\end{enumerate}

\end{enumerate}
Here $\underline{\mathbf I}(\mathcal R)$ and  $\overline{\mathbf I}(\mathcal R)$ denote the lower and upper bound values of feature $\mathbf I$, respectively. Let $\mathbf Z=\left(\mathbf X,\mathbf y \right)$ denote the data set, where $\mathbf X$ and $\mathbf y$ denote the feature matrix and the vector of labels, respectively,  then the lower and upper bound values of  feature $\mathbf I$  of the entire feature space are  as follows. 
\begin{align*}
\mbox{~Upper Bound~ :}&\overline{\mathbf I}=\max_{\mathbf x \in \mathbf X} \mathbf x_{\mathbf I}\\
\mbox{~Lower Bound~ :}&\underline{\mathbf I}=\min_{\mathbf x \in \mathbf X} \mathbf x_{\mathbf I}  \end{align*}

\subsection{Tree Construction}
In this paper, we consider the classification and regression  tree (CART)~\cite{Breiman:2253780} for feature space partitioning. CART uses  Gini index to measure the purity of a  feature space partition.  Gini index is a measure of how often a randomly chosen element from the set would be incorrectly labeled if it was randomly labeled according to the distribution of labels in the subset.   Suppose $|\mathcal R^k|$ and $|\mathcal R|$ denote the weighted number of data points labeled $k$ and the weighted number of all the data points in  $\mathcal R$, respectively. Then the Gini index of $\mathcal R$ can be computed as follows.
\begin{equation}
G(\mathcal R)=\sum_{k=1}^K \frac{|\mathcal R^k|}{|\mathcal R|}\left (1-\frac{|\mathcal R^k|}{|\mathcal R|}  \right)
\end{equation}

Suppose feature space  $\mathcal R$  is split into  $\mathcal R_1$ and $\mathcal R_2$ by feature  $\mathbf I$ and value $s$. CART will always select the split point to minimize the weighted Gini index of  $\mathcal R_1$ and $\mathcal R_2$.
\begin{equation}
\label{eq:mingini}
\langle \mathbf I,~s \rangle  =\arg \min_{\langle \mathbf I,~s \rangle} \hat{G}(\mathcal R, \mathbf I,s), \mbox{ where} 
\end{equation}
$$
\hat{G}(\mathcal R, \mathbf I,s)=\frac{|\mathcal R_1|}{|\mathcal R|} G(\mathcal R_1)+ \frac{|\mathcal R_2|}{|\mathcal R|}G(\mathcal R_2)\\
$$
The  gain in the Gini index from the split is then 
\begin{equation}
\Delta G(\mathcal R)=G(\mathcal R)-\min_{\mathbf I,~s}\hat{G}(\mathcal R, \mathbf I,s)
\end{equation}

\my{Let $+$ and $-$ denote the label of training data points and  E-points, respectively. }
Suppose there are infinite  E-points  uniformly distributed in the hyper-rectangle $\mathcal R$. If we set the weight of each E-point  to $\frac{|\mathcal R^-|}{\infty}$,  the weighted number of E-points\footnote{In the rest of this paper, when we refer to the number of E-points, what we actually mean is the weighted number.} in $\mathcal R$  is equal to  $\frac{|\mathcal R^-|}{\infty}\times \infty =|\mathcal R^-|$.  \my{Let $|\mathcal R^+|$ denote the weighted number\footnote{We assume the weight of a single training sample is equal to $1$.} of training points in the hyper-rectangle $\mathcal R$.} The weighted Gini index after $\mathcal R$ is split  into  $\mathcal R_1$ and $\mathcal R_2$ is equal to 
\begin{align}
\label{eq:e0}
\frac{|\mathcal R_1^+|+|\mathcal R_1^-|}{|\mathcal R^+|+|\mathcal R^-|}&\times G(\mathcal R_1)+ \frac{|\mathcal R_2^+|+|\mathcal R_2^-|}{|\mathcal R^+|+|\mathcal R^-|}\times G(\mathcal R_2)~\mbox{, where}\\
|\mathcal R_1^-|&=\frac{s-\underline{\mathbf I}(\mathcal R)}{\overline{\mathbf I}(\mathcal R)-\underline{\mathbf I}(\mathcal R)}\times |\mathcal R^-|=\frac{s-\underline{\mathbf I}(\mathcal R)}{\Delta{\mathbf I}(\mathcal R)}\times |\mathcal R^-|
\label{eq:e1}\\
|\mathcal R_2^-|&=\frac{\overline{\mathbf I}(\mathcal R)-s}{\overline{\mathbf I}(\mathcal R)-\underline{\mathbf I}(\mathcal R)}\times |\mathcal R^-|=\frac{\overline{\mathbf I}(\mathcal R)-s}{\Delta{\mathbf I}(\mathcal R)}\times |\mathcal R^-|
\label{eq:e2}
\end{align}

\section{CART Customizations for FSPT Feasibility}
\label{sec:cust}
In this section, we propose several customization techniques for CART to realize FSPT so that it is suitable for identifying feature space partitions without sufficient training samples.

\subsection{Stopping Criterion and the Number of  E-points}
\label{sec:stop}
\begin{figure}[t]
\centering
\includegraphics[width=0.25\textwidth]{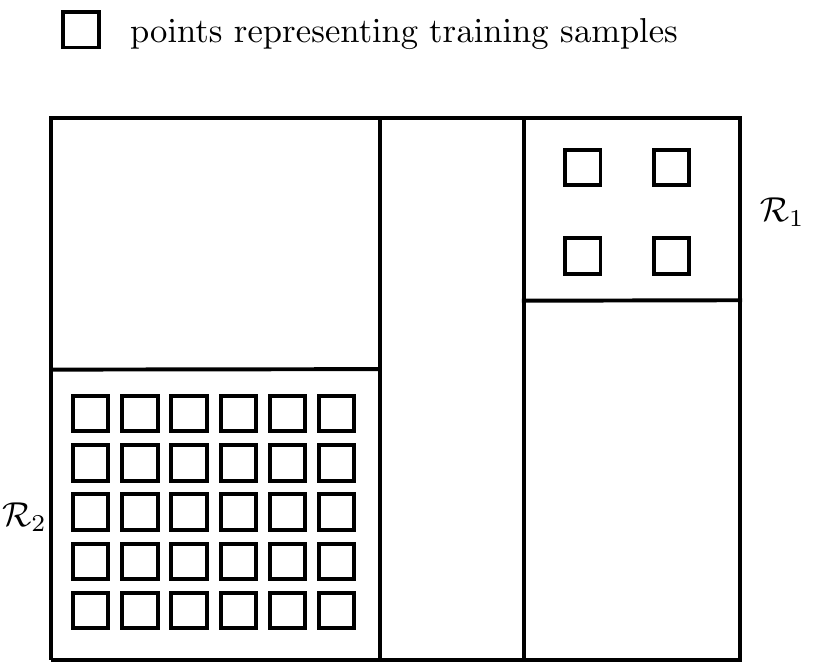}
\caption{Stop Construction}
\label{fig:stop}
\end{figure}

The first question we must answer is when can we \emph{stop constructing the tree} because otherwise, the tree can grow infinitely.  In our case, we can stop further splitting feature space partition $\mathcal R$  if all training samples are  uniformly distributed in $\mathcal R$ approximately. As shown  in Figure~\ref{fig:stop}, we can stop splitting $\mathcal R_1$ and $\mathcal R_2$  because data points are evenly distributed in them, even though  the densities of training samples  are different.  The intuition for  the stopping criterion is that, when the stopping condition is satisfied, any further split can only generate two hyper-rectangles with similar data distributions. 

Another parameter we need to determine is the number of E-points $|\mathcal R^-|$. According to  Equations~\ref{eq:e1} and \ref{eq:e2}, the number of E-points $|\mathcal R^-|$for hyper-rectangle $\mathcal R$ could decrease drastically if the split value $s$ is close to the bound of feature $\mathbf I$, i.e., $\overline{\mathbf I}(\mathcal R)-s\rightarrow 0$ or $s-\underline{\mathbf I}(\mathcal R)\rightarrow 0$. For a high dimension space, $|\mathcal R^-|$ will become close to 0 after several splits.  As a result,  any further division can only result in a negligible decrease in the Gini Index \my{and hence a negligible increase in the gain of Gini Index}, which makes it more  difficult to find the optimal split point.

\begin{figure}[t]
\centering
\includegraphics[width=0.35\textwidth]{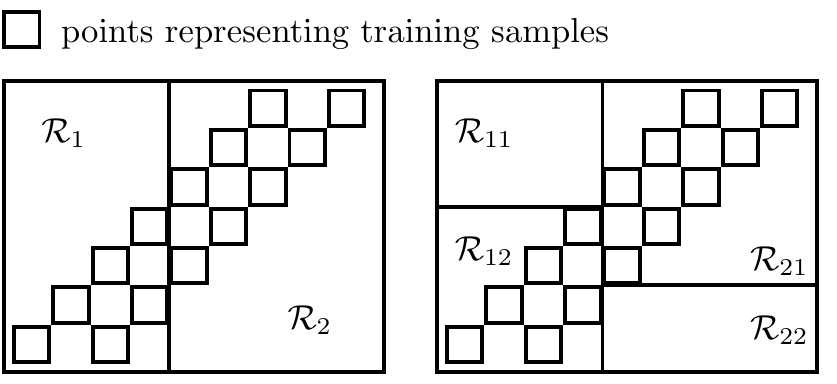}
\caption{Exceptional Scenario}
\label{fig:stop2}
\end{figure}

To address the above two issues, we fix $|\mathcal R^-|=|\mathcal R^+|$ at each split.  In this case ($|\mathcal R^+|= |\mathcal R^-|$), when the condition, i.e.,  all training samples are uniformly distributed in  $\mathcal R$ approximately, is satisfied,    the gain in Gini index  $\Delta G(\mathcal R)$ will be close to $0$.
$$
\overbrace{G(\mathcal R)}^{=0.5}-\frac{|\mathcal R_1|}{|\mathcal R|}\times \overbrace{G(\mathcal R_1)}^{\rightarrow 0.5}-\frac{|\mathcal R_2|}{|\mathcal R|}\overbrace{G(\mathcal R_2)}^{\rightarrow 0.5}\rightarrow 0
$$
Thus, the gain in Gini index  $\Delta G(\mathcal R)\rightarrow 0 $  is a necessary  condition  to indicate whether the stopping criterion (training samples are evenly distributed in $\mathcal R$ approximately) is satisfied. However, it is not a sufficient condition.  Figure~\ref{fig:stop2} shows an exceptional scenario  where $\Delta G(\mathcal R)$ is close to $0$, but training samples are not distributed uniformly in $\mathcal R$.  

To address this issue,  we use a counter $c$ to record the number of  successive times that  $\Delta G(\mathcal R)\leq \epsilon $ where $\epsilon\rightarrow 0$.  The  construction process  terminates  when $c$ is greater than some threshold $\lambda$.  For example,  in Figure~\ref{fig:stop2}, $ \Delta G(\mathcal R)\leq \epsilon$ and hence $c\leftarrow c+1$. After we split $\mathcal R$  into $\mathcal R_1$ and $\mathcal R_2$, 
\begin{align*}
 \Delta G(\mathcal R_1)=G(\mathcal R_1)-\frac{|\mathcal R_{11}|}{|\mathcal R_1|}G(\mathcal R_{11})-\frac{|\mathcal R_{12}|}{|\mathcal R_1|}G(\mathcal R_{12})>\epsilon\\   
\Delta G(\mathcal R_2)=G(\mathcal R_2)-\frac{|\mathcal R_{21}|}{|\mathcal R_2|}G(\mathcal R_{21})-\frac{|\mathcal R_{22}|}{|\mathcal R_2|}G(\mathcal R_{22})>\epsilon   
\end{align*}
and  hence the counter c is reset to $0$. As a result, the tree construction algorithm will continue to  split  $\mathcal R_1\rightarrow \{\mathcal R_{11},\mathcal R_{12}\}$  and $\mathcal R_2\rightarrow \{\mathcal R_{21},\mathcal R_{22}\}$.

Meanwhile, $\lambda$ need not necessarily be a fixed value. When there are lots of training samples in  $\mathcal R$, then we can assign a larger value to $\lambda$. Otherwise, if  $|\mathcal R^+|$  is very small, the construction process can terminate immediately. 


\subsection{Split Points}
\label{sec:split}
One of the critical problems in tree learning is to find the best split as represented by Equation~\ref{eq:mingini}. A simple greedy solution is to  enumerate over all possible split values on all the features.   In our case, however, there are infinite possible  split values, because   we  assume there are infinite E-points  uniformly distributed in the feature space.

Suppose we want to find the best split for the hyper-rectangle $\mathcal R$ on feature $\mathbf I$.  Let  $\mathcal S_{\mathbf I}=\{ s_{1},  s_{2},\ldots,s_N\}$ denote the set of unique values of feature $\mathbf I$ of training samples in $\mathcal R$, where items in $\mathcal S_{\mathbf I}$ are sorted in ascending order, i.e., $\underline{\mathbf I}(\mathcal R) \leq  s_1  < \ldots <  s_{N}\leq  \overline{\mathbf I}(\mathcal R)$.  Let $\mbox{cdf}(s)$ denote the cumulative distribution function of \emph{training samples} in  $\mathcal R$  on feature  $\mathbf I$. Then $\forall s\in [s_k,s_{k+1})$, we have $\mbox{cdf}(s)=\mbox{cdf}(s_k)$ \my{because there exists no training sample with feature $\mathbf I$ within $(s_k,s_{k+1})$}.  For simplicity, we assume feature $\mathbf I$ is normalized to 1, i.e.,  $\Delta \mathbf I(\mathcal R)=\overline{\mathbf I}(\mathcal R)-\underline{\mathbf I}(\mathcal R)=1$, and hence the split value $s\in[0,1]$. If we split  $\mathcal R$ into $\mathcal R_1$  and $\mathcal R_2$ by value $s$ and  feature $\mathbf I$, then

\begin{align*}
 &G(\mathcal R_1)=2\times \left( \frac{|\mathcal R_1^+|}{|\mathcal R_1^+|+|\mathcal R_1^-|}\right)\left(\frac{|\mathcal R_1^-|}{|\mathcal R_1^+|+|\mathcal R_1^-|}\right)\\
 &=2\left(\frac{\mbox{cdf}(s)}{\mbox{cdf}(s)+s}\right)\left(\frac{s}{\mbox{cdf}(s)+s}\right)\\
&G(\mathcal R_2)=2\times \left( \frac{|\mathcal R_2^+|}{|\mathcal R_2^+|+|\mathcal R_2^-|}\right)\left(\frac{|\mathcal R_2^-|}{|\mathcal R_2^+|+|\mathcal R_2^-|}\right)\\
&=2\left(\frac{1-\mbox{cdf}(s)}{2-\mbox{cdf}(s)-s}\right)\left(\frac{1-s}{2-\mbox{cdf}(s)-s}\right)\\
&\Rightarrow \hat{G}(\mathcal R, \mathbf I,s)= \frac{\mbox{cdf}(s)+s}{2}\times G(\mathcal R_1)+ \frac{2-s-\mbox{cdf}(s)}{2}\times G(\mathcal R_2)\\
&= \left(\frac{\mbox{cdf}(s)s}{\mbox{cdf}(s)+s}\right)+\left(\frac{(1-\mbox{cdf}(s))(1-s)}{2-\mbox{cdf}(s)-s}\right)
\end{align*}
Let's compute the partial derivative of weighted Gini index on $s$.
\begin{align*}
&\frac{\partial \hat{G}(\mathcal R, \mathbf I,s)}{\partial s}=\left(\frac{\mbox{cdf}(s)}{\mbox{cdf}(s)+s}\right)^2-\left(\frac{1-\mbox{cdf}(s)}{2-\mbox{cdf}(s)-s}\right)^2=\\
&\overbrace{\left(\!\!\frac{\mbox{cdf}(s)}{\mbox{cdf}(s)+s}\!+\!\frac{1-\mbox{cdf}(s)}{2-\mbox{cdf}(s)-s}\!\!\right)}^{>0} \!\!\left(\!\!\frac{\mbox{cdf}(s)-s}{(\mbox{cdf}(s)+s)(2-\mbox{cdf}(s)-s)}\!\!\right)
\end{align*}
When $s\in [s_k,s_{k+1})$,  $\mbox{cdf}(s)=\mbox{cdf}(s_k)$ is a constant value.   If $s_k\geq \mbox{cdf}(s_k)\Rightarrow \frac{\partial \hat{G}(\mathcal R, \mathbf I,s)}{\partial s}\leq 0$, then $\hat{G}(\mathcal R, \mathbf I,s)$ is minimized when $s=s_{k+1}-\epsilon$,  where $\epsilon\rightarrow +0$.  If $s_{k+1}\leq \mbox{cdf}(s_k)\Rightarrow \frac{\partial \hat{G}(\mathcal R, \mathbf I,s)}{\partial s}\geq 0$, then $\hat{G}(\mathcal R, \mathbf I,s)$ is minimized when $s=s_{k}$.  Finally if  $s_{k}< \mbox{cdf}(s_k)<s_{k+1}$, then  $\hat{G}(\mathcal R, \mathbf I,s)$ is minimized when $s=s_{k}$  or  $s=s_{k+1}-\epsilon$.

Therefore, there is no need to try all infinite split values, and the potential split value set is reduced to $$\left\{\max\left(\underline{\mathbf I}(\mathcal R),s_1-\epsilon\right),s_{1},s_{2}-\epsilon,s_2,s_3-\epsilon,\ldots, s_{N}\right \}$$ It can still be  computationally demanding  to  enumerate all the split points in  the candidate set, especially when the data cannot fit entirely into memory. To address this issue, we can also use  approximate split finding algorithms (e.g., use candidate splitting points according to percentiles of feature distribution), which is quite common in tree learning algorithms.



\subsection{New Splitting Criterion and Score Function}
In general, more training samples are required for an ML model to achieve good performance within a feature space partition $\mathcal R$ with a larger volume $V(\mathcal R)$, where 
\begin{align*}
V(\mathcal R)=    \prod_{\mathbf I} \left(\overline{\mathbf I}(\mathcal R)-\underline{\mathbf I}(\mathcal R) \right)=    \prod_{\mathbf I} \Delta \mathbf I(\mathcal R)\\
\end{align*}
This scenario is clearly shown in Example~\ref{example:toy2} in Section~\ref{sec:moti}. While two feature space partitions  $x\in [0,5]$ and  $x\in (10,15]$ have the same volume, the ML model fits the sine function much better  in the former  with 40 training samples than in the latter with 10 training samples.   Of course, this is not always the case, especially when the model is poorly trained. For example, imagine we have a model whose output is always zero. Then it is evident that its performance has a very weak dependence on the training samples.   Thus,  the premise to infer ML models'  performance  in different feature space partitions from training samples and volume $V(\mathcal R)$ is that it does not over-fit or under-fit.

Another challenge in inferring ML models'  performance in $\mathcal R$  from its volume and training samples is that all features are equally important in the computation of  volume.  Besides, as long as there exists any feature $\mathbf I$ with $ \Delta \mathbf I(\mathcal R)$ close to 0, the volume of $\mathcal R$ will also be close to 0 irrespective of the feature importance of $\mathbf I$ and the side length of other features.

\begin{figure}
\centering
\begin{minipage}[t]{0.2\textwidth}
  \centering
  \includegraphics[width=\textwidth]{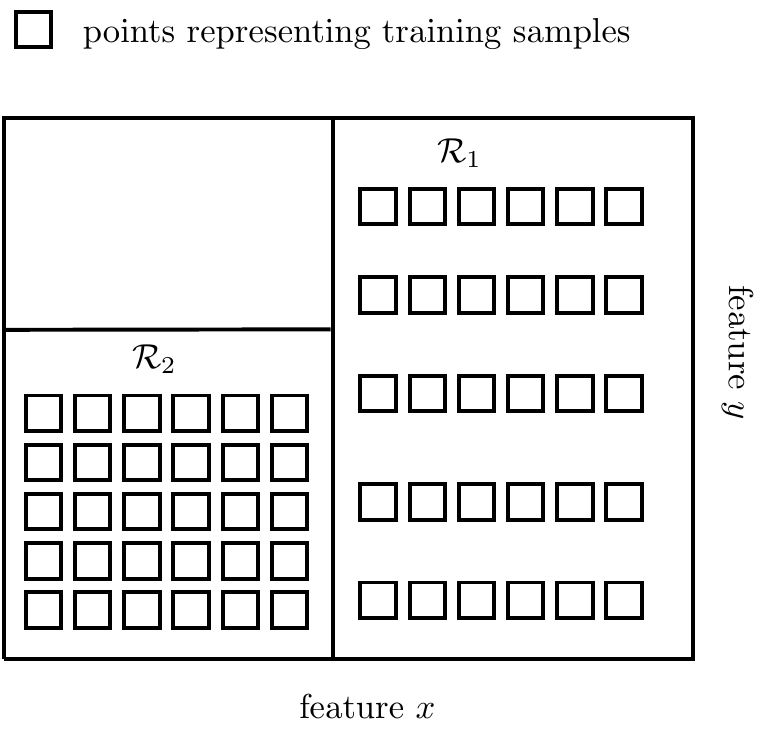}
\caption{Problem of Irrelevant Features}
\label{fig:importance}
\end{minipage}
\begin{minipage}[t]{0.18\textwidth}
  \centering
  \includegraphics[width=\textwidth]{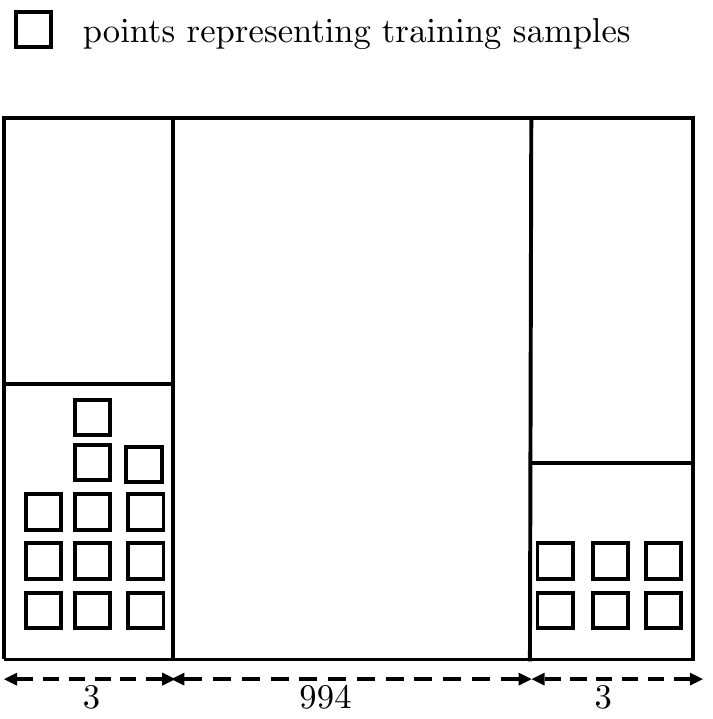}
\caption{Resulting hyper-rectangles have too small rectangles}
\label{fig:small}
\end{minipage}
\end{figure}

As shown in Figure~\ref{fig:importance}, $\mathcal R_1$  and $\mathcal R_2$ have the same number of training samples, but $\mathcal R_1$  covers a larger feature space.   If feature $x$ and feature $y$ are equally important, then we can expect that  an ML model will have a higher fitting degree of the objective function in $\mathcal R_2$  than in $\mathcal R_1$, if it does not over-fit or under-fit.  However, if feature $y$ is of very low importance or irrelevant to the objective function, then the previous inference can be misleading.

To address this problem, we can exclude all the irrelevant features, and  meanwhile,  incorporate feature importance into the splitting criterion as well as the score function for  the resulting feature space partitions.  There are lots of techniques that can be used to  assess feature importance~\cite{gevrey2003review,williamson2017nonparametric,breiman2001random}.  Of course, it is not trivial  to obtain precise feature importance values. Meanwhile, features that are globally important may not be important in the local context, and vice versa. In this paper, we only consider global feature importance with constant values.

Another issue is that we also do not want to split the feature space into too many hyper-rectangles with an extremely small  side length on a particular feature $\mathbf I$,  i.e.,  $\frac{\Delta \mathbf I(\mathcal R)}{\Delta \mathbf I}\rightarrow 0$, where $\Delta \mathbf I$ is the side length of the entire feature space.   For example, in Figure~\ref{fig:small}, since  $\frac{\Delta \mathbf I(\mathcal R)}{\Delta \mathbf I}=0.3\%\rightarrow 0$, we will not further split on this feature unless it can bring significant gain in the splitting criterion. 

To meet the requirements mentioned  above,  we propose:
\begin{enumerate}
    \item  A more reasonable splitting criterion  named weighed gain in Gini index, which considers  1) $\hat{G}(\mathcal R, \mathbf I,s)$, 2) the feature importance values $\{f_1,f_2,\ldots,f_d\}$ and 3) the side length of each feature $\Delta \mathbf I(\mathcal R)$.
    \item A heuristic score function to assess the resulting feature space partitions.
\end{enumerate}
\subsubsection{Splitting Criterion}
\label{sec:newsplit}

Equation~\ref{eq:gini_feature} shows the new criterion for selecting splitting points.
\begin{equation}
\label{eq:gini_feature}
\langle \mathbf I,~s \rangle  =\arg \max_{\langle \mathbf I,~s \rangle}\frac{ f_{\mathbf I}\times \Delta \mathbf I(\mathcal R)}{\Delta \mathbf I}\Delta G(\mathcal R)\\
\end{equation}
$$
\mbox{where  }\Delta G(\mathcal R)=\left(G(\mathcal R)-\hat{G}(\mathcal R, \mathbf I,s)\right)
$$
The intuition of this criterion is simple. We prefer to split on  features that are important and  have a larger side length, unless there is a significant increase in $\Delta G(\mathcal R)$.  



\subsubsection{Score Function}
\label{sec:score}
We propose a heuristic score function in Equation~\ref{eq:score} to evaluate the resulting hyper-rectangle $\mathcal R$,   and  it takes both the feature  importance  and  the side length of each feature into consideration.
\begin{equation}
\label{eq:score}
S(\mathcal R)= \sum_{\mathbf I}f_{\mathbf I}\times \frac{|\mathcal R^+|}{|\mathcal R^+|+\frac{\Delta \mathbf I(\mathcal R)}{\Delta \mathbf I}\times E}
\end{equation}
In Equation~\ref{eq:score},  $E$ is a hyperparameter. We set $E=\frac{N}{d}$ in our experiments in Section~\ref{sec:evaluation}, where $N$ is the number of training samples and $d$ is the number  of features.    The intuition for Equation~\ref{eq:score} is that we evaluate $\mathcal R$ on each feature separately based the training samples and side length $\Delta \mathbf I(\mathcal R)$. For example, if    $ \Delta \mathbf I(\mathcal R)\rightarrow 0$, i.e., training samples in $\mathcal R$ have the same value on feature $\mathbf I$, then  $\mathcal R$ is supposed to get the full score   of feature $\mathbf I$ (i.e., $f_{\mathbf I}$). After FSPT completes the tree construction, the final  scores for each partition can be normalized into the range $[0,1]$. 

\noindent \textbf{Complexity}: \my{Suppose FSPT has depth $T$, then the score of an input instance can be calculated efficiently with complexity $O(T)$.  Since each feature space partition contains at least one training sample,  the run-time complexity is $O(\log N)$

}

\section{Reject Model}
\label{sec:rej}
Suppose input instance $x$ is in feature space partition $\mathcal R$, the output of FSPT for $x$ is  $$\phi_{F}(x)=S(\mathcal R)$$

\noindent\textbf{Regression Problems}:  We can use $\phi_{F}(x)$ to determine whether the prediction for input instance $x$ should be rejected in regression problems directly. Let $\phi_{M}(x)$ denote the predictive output of ML models for input instance $x$, then a reject option can be selected when $\phi_{F}(x)$ is smaller than a particular threshold $t$. 
\begin{align}
\label{eq:reject1}
\hat y(x)=\begin{cases}\phi_{M}(x)~\mbox{ if  }\phi_{F}(x)\geq t\\\mbox{reject option, otherwise }\end{cases}
\end{align}
\noindent\textbf{Classification Problems}: Suppose $\phi_{M}^c(x)$ denotes  ML models' predictive probability that $x$ belongs to  class  ``c'', then  the final predictive class for  $x$ is $$c=\arg \max_c \{\phi_{M}^c(x)\}$$ Let $\phi_{M}(x)=\max_c \{\phi_{M}^c(x)\}$,  we  use both $\phi_{M}(x)$  and $\phi_{F}(x)$  to determine whether a reject option should be selected. The reason to adopt such a reject strategy is that an  ML model's accuracy for an input instance  depends on its distance to the decision boundaries significantly.  

\begin{figure}[h!]
\centering
\includegraphics[width=0.22\textwidth]{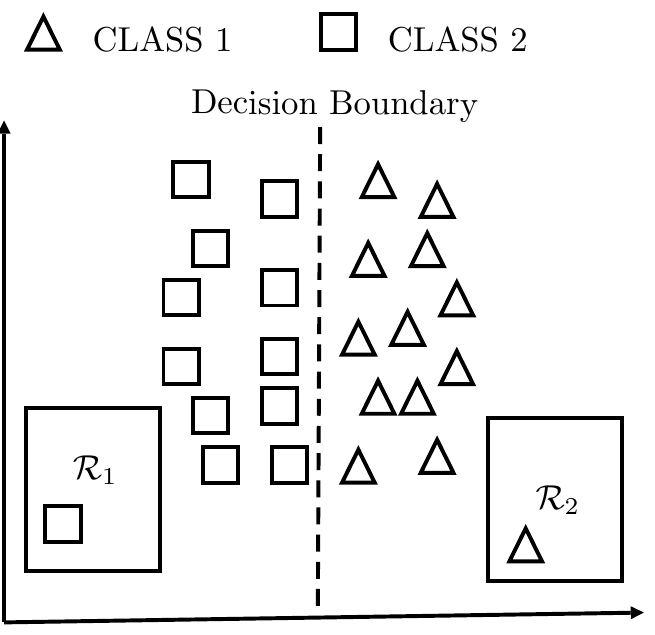}
\caption{Training samples for a binary classification problem}
\label{fig:toy3}
\end{figure}

Figure~\ref{fig:toy3}  shows a toy example of a binary classification problem. As we can see,  few training samples exist in feature space partition $\mathcal  R_1$ and $\mathcal  R_2$ perhaps due to small probability density there. Thus, FSPT gives relatively low scores to $\mathcal  R_1$ and $\mathcal  R_2$ because there is not sufficient evidence to draw a confident conclusion. 
However,  the model may have  $100\%$ accuracy in $\mathcal  R_1$ and $\mathcal  R_2$ just because its  guess    is ``lucky'' enough  to  be  correct. 

Thus, for classification problems, the output of FSPT will be used as a complement to the ML model, and the prediction of ML models will be  adopted only when both $\phi_{F}(x)$  and  $\phi_{M}^c(x)$ exceed the corresponding thresholds. 
\begin{align}
\label{eq:reject2}
\hat y(x)=\begin{cases} c~\mbox{ if  }\phi_{M}^c(x)\geq t_1\wedge \phi_{F}(x)\geq t_2 \\\mbox{reject option, otherwise }\end{cases}
\end{align}

\section{Evaluation}
\label{sec:evaluation}
\begin{figure}[h!]
\centering
\includegraphics[width=0.375\textwidth]{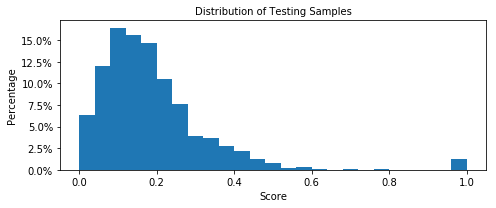}
\caption{Mining Process:   sample distribution}
\label{fig:mining_dist}
\end{figure}

\begin{figure}[h!]
\centering
\includegraphics[width=0.375\textwidth]{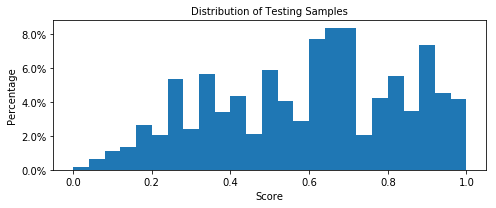}
\caption{SCITOS:    sample distribution}
\label{fig:scitosdist}
\end{figure}

\begin{figure}[h!]
\centering
\includegraphics[width=0.375\textwidth]{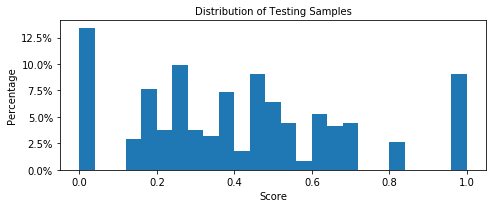}
\caption{Breast Cancer Diagnosis:    sample distribution}
\label{fig:breastdist}
\end{figure}

In this section, we evaluate the effectiveness of our proposed technique  for both regression and classification problems.   The feature importance values used in the experiments are obtained from  Random Forest~\cite{breiman2001random}. We investigate  the performance of three popular ML models, i.e.,  
\begin{enumerate}
  \item Neural Networks (NN)
  \item Support Vector Machine (SVM)
  \item Gaussian Process (GP)
\end{enumerate}
when we set different rejection thresholds for them.  Since GP can offer standard deviations for testing samples in regression problems, we also investigate the relationship between the  standard deviations  and  $\phi_{F}(x)$. For a given data set  $\mathbf Z$, testing samples are  randomly sampled from $\mathbf Z$.  \my{In Figure~\ref{fig:mining_dist}, \ref{fig:sarcosdist}, \ref{fig:scitosdist} and \ref{fig:breastdist}, we show the distributions of testing samples  with respect to the scores.}


\subsection{Regression Problems}
\begin{figure}[h!]
\centering
\includegraphics[width=0.44\textwidth]{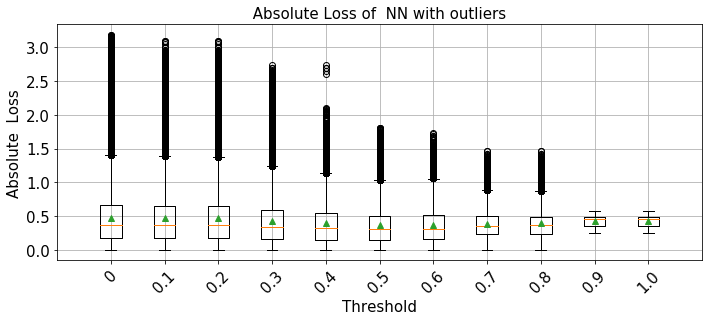}
\caption{Mining Process:    absolute loss for NNs}
\label{fig:mining_mlpbox}
\end{figure}

\begin{figure}[h!]
\centering
\includegraphics[width=0.44\textwidth]{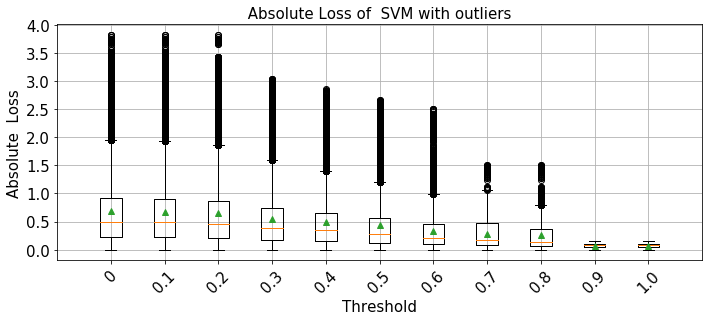}
\caption{Mining Process:  absolute loss for SVM}
\label{fig:mining_svmbox}
\end{figure}

\begin{figure}[h!]
\centering
\includegraphics[width=0.44\textwidth]{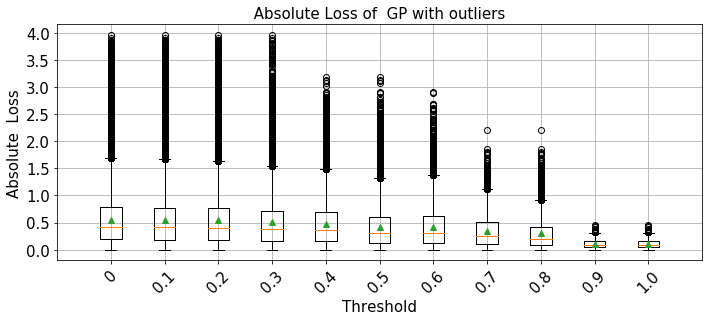}
\caption{Mining Process: absolute loss for GP}
\label{fig:mining_gpbox}
\end{figure}

\begin{figure}[h!]
\centering
\includegraphics[width=0.44\textwidth]{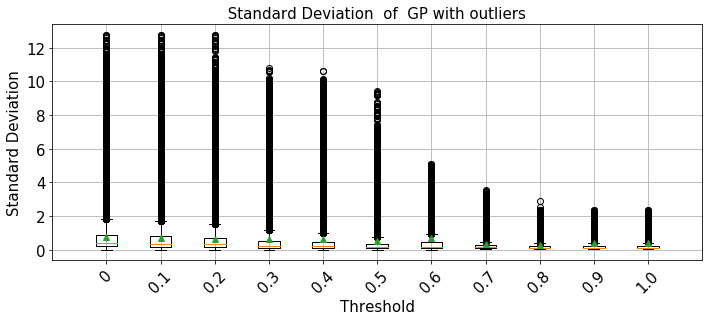}
\caption{Mining Process: standard deviation for GP}
\label{fig:mining_gpx2}
\end{figure}

\begin{figure}[h!]
\centering
\includegraphics[width=0.375\textwidth]{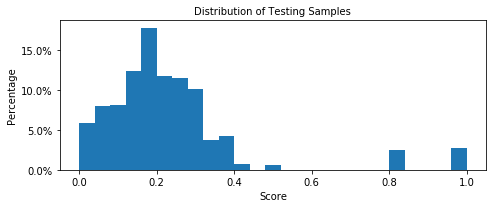}
\caption{SARCO:    sample distribution}
\label{fig:sarcosdist}
\end{figure}

\begin{figure}[h!]
\centering
\includegraphics[width=0.44\textwidth]{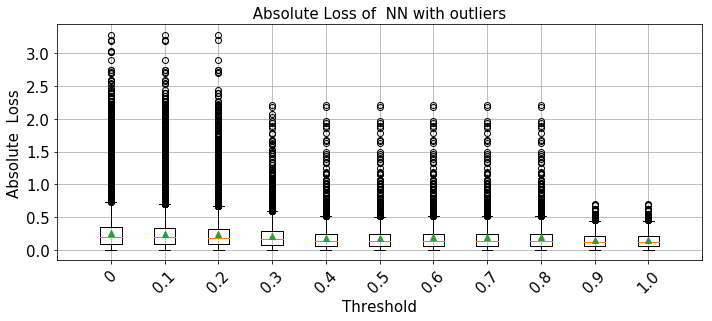}
\caption{SARCO: absolute loss for NNs}
\label{fig:sarcos_mlpbox}
\end{figure}

\begin{figure}[h!]
\centering
\includegraphics[width=0.44\textwidth]{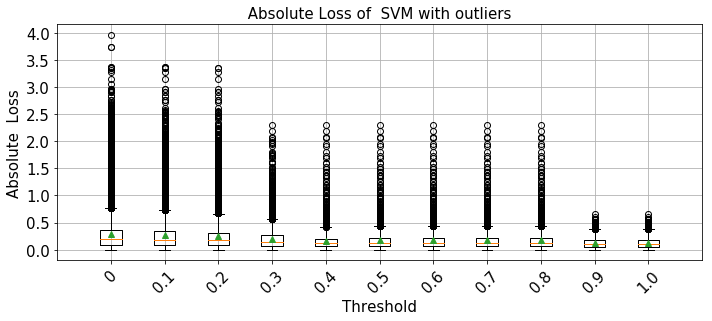}
\caption{SARCO: absolute loss for SVM}
\label{fig:sarcos_svmbox}
\end{figure}

\begin{figure}[h!]
\centering
\includegraphics[width=0.44\textwidth]{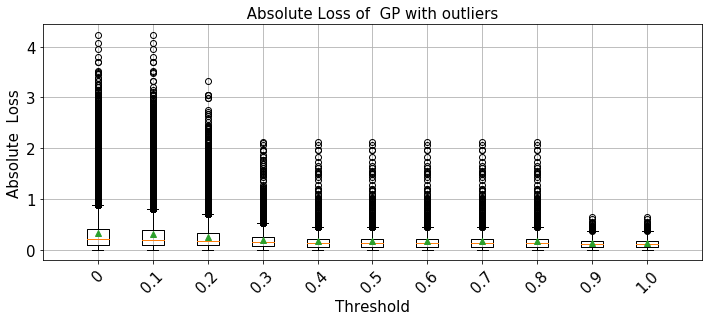}
\caption{SARCO: absolute loss for GP}
\label{fig:sarcos_gpbox}
\end{figure}

\begin{figure}[h!]
\centering
\includegraphics[width=0.44\textwidth]{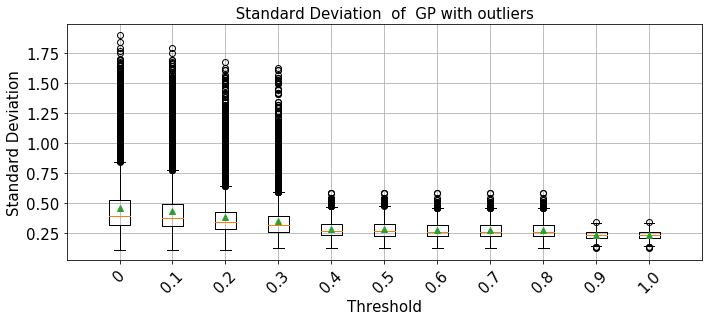}
\caption{SARCO: standard deviation for GP}
\label{fig:sarcos_gpx2}
\end{figure}

For regression problems,  we evaluate the reject model in Equation~\ref{eq:reject1}.   We use box-plot (Figure~\ref{fig:mining_mlpbox} to \ref{fig:sarcos_gpx2}) to show absolute loss $|y-\hat y|$/standard deviation of  testing samples with different rejection thresholds $t\in \{0,0.1,0.2,0.3,\ldots,1\}$.  We consider  the following two applications in the experiments.    


\noindent \textbf{Quality Prediction in a Mining Process}:  This dataset  is about a flotation plant which is a process used to concentrate the iron ore~\cite{mining_process}.  The goal of this task is to predict  the percentage of Silica at the end of the process from 22 features. As this value is measured every hour, if we can predict how much silica is in the ore concentrate, we can help the engineers, giving them early information to take actions. Hence, they will be able to take corrective actions in advance and also help the environment by reducing the amount of ore that goes to tailings.

\noindent \textbf{The Inverse Dynamics of a  SARCOS Robot Arm}:  This task is to map from a 21-dimensional input space (7 joint positions, 7 joint velocities, 7 joint accelerations) from a seven degrees-of-freedom SARCOS robot arm~\cite{SARCOS} to the  inverse dynamics of a corresponding torque.  An inverse dynamics model can be used in the following manner: a planning module decides on a trajectory that takes the robot from its start to goal states, and this specifies the desired positions, velocities and accelerations at each time. The inverse dynamics model is used to compute the torques needed to achieve this trajectory and errors are corrected using a feedback controller~\cite{seeger2004gaussian}.

\subsection{Classification Problems}
 For classification problems, we evaluate the reject model in Equation~\ref{eq:reject2}.  Through the experiment results, we show that input instance with both high scores from  FSPT $\phi_F(x)$  and  high scores (i.e., predictive probability output  $\phi_M(x)$) from  ML models has  a lower error rate.  Let  $t_1$ denote the rejection threshold for  ML model, and $t_2$ denote rejection threshold for FSPT. We partition testing samples into group $G_1$ and $G_2$.
\begin{enumerate}
    \item $G_1$: $ \phi_M(x)\geq t_1\wedge  \phi_F(x) \geq t_2$
    \item $G_2$: $ \phi_M(x) \geq t_1 \wedge \phi_F(x) < t_2$
\end{enumerate}
From Table~\ref{fig:scitos_nn} to Table~\ref{fig:bcd_gp}, we present the accuracy of $G_1$ and $G_2$ for different combinations of  $t_1$ and  $t_2$. Besides, we also show the mean value of $\phi_M(x)$ and proportion of  $G_1$ and $G_2$. We show that even when  $G_1$ and $G_2$ have similar mean values of $\phi_M(x)$, which indicates that ML models have similar confidence in the predictions of the testing samples in  $G_1$ and $G_2$, their accuracy of $G_1$ is  higher than $G_2$.  We consider  the following two applications in the experiments.

\noindent \textbf{The Navigation Task for Mobile Robot SCITOS-G5}:  This task is to map 24-dimension input, i.e., sensor readings  from the 24 sensors of  the mobile robot SCITOS-G5~\cite{Dua:2017} to 4 classes of behaviors: 1) ``Move-Forward'', 2) ``Sharp-Right-Turn'', 3)  ``Slight-Left-Turn'' and 4)  ``Slight-Right-Turn''.      ML models will be  trained for the  mobile robot  to take decisions that determine its correct movement.    

\begin{table}[h]
\caption{SCITOS: NNs}
\label{fig:scitos_nn}
\scriptsize
\center
\begin{tabular}{|l|l|l|l|l|l|l|l|l|}
  \hline
\multicolumn{1}{c}{ $t_1~t_2$}&\multicolumn{3}{c}{$G_1$} &  \multicolumn{3}{c}{$G_2$}\\
  \hline
&  Proportion & Accuracy  &$\phi_M(x)$    & Proportion&  Accuracy &  $\phi_M(x)$  \\
0.90 0.00 & 1.00  & 0.925& 0.970 & 0.00 & nan   &nan  \\
0.90 0.30 & 0.86  & 0.925& 0.970 & 0.14 & 0.921 &0.972  \\
0.90 0.60 & 0.55  & 0.930& 0.970 & 0.45 & 0.918 &0.970  \\
0.90 0.90 & 0.14  & 0.931& 0.972 & 0.86 & 0.924 &0.970  \\
0.95 0.00 & 1.00  & 0.947& 0.984 & 0.00 & nan   &nan  \\
0.95 0.30 & 0.85  & 0.948& 0.984 & 0.15 & 0.944 &0.984  \\
0.95 0.60 & 0.54  & 0.951& 0.984 & 0.46 & 0.942 &0.984  \\
0.95 0.90 & 0.15  & 0.956& 0.984 & 0.85 & 0.946 &0.984  \\
 \hline
\end{tabular}
\end{table}

\begin{table}[h]
\caption{SCITOS: SVM}
\label{fig:scitos_svm}
\scriptsize
\center
\begin{tabular}{|l|l|l|l|l|l|l|l|l|}
  \hline
\multicolumn{1}{c}{ $t_1~t_2$}&\multicolumn{3}{c}{$G_1$} &  \multicolumn{3}{c}{$G_2$}\\
  \hline
&  Proportion & Accuracy  &  $\phi_M(x)$   & Proportion&  Accuracy &$\phi_M(x)$  \\
0.90 0.00&1.00 & 0.960 & 0.963 &0.00& nan  & nan  \\
0.90 0.30&0.86 & 0.961 & 0.963 &0.14& 0.954& 0.962\\  
0.90 0.60&0.56 & 0.967 & 0.962 &0.44& 0.950& 0.963\\  
0.90 0.90&0.15 & 0.959 & 0.961 &0.85& 0.960& 0.963\\  
0.95 0.00&1.00 & 0.968 & 0.979 &0.00& nan  & nan  \\
0.95 0.30&0.86 & 0.970 & 0.979 &0.14& 0.958& 0.979\\  
0.95 0.60&0.56 & 0.975 & 0.978 &0.44& 0.959& 0.980\\  
0.95 0.90&0.14 & 0.974 & 0.978 &0.86& 0.967& 0.979\\ 
 \hline
\end{tabular}
\end{table}

\begin{table}[h]
\caption{SCITOS:  GP}
\label{fig:scitos_gp}
\scriptsize
\center
\begin{tabular}{|l|l|l|l|l|l|l|l|l|}
  \hline
\multicolumn{1}{c}{ $t_1~t_2$}&\multicolumn{3}{c}{$G_1$} &  \multicolumn{3}{c}{$G_2$}\\
  \hline
&  Proportion & Accuracy  &  $\phi_M(x)$  & Proportion&  Accuracy &$\phi_M(x)$ \\
0.60 0.00& 1.00 & 0.975& 0.703&  0.00 & nan  & nan  \\
0.60 0.30& 0.52 & 0.989& 0.712&  0.48 & 0.960& 0.694  \\
0.60 0.60& 0.12 & 1.000& 0.738&  0.88 & 0.972& 0.698  \\
0.60 0.90& 0.02 & 1.000& 0.782&  0.98 & 0.975& 0.702  \\
0.70 0.00& 1.00 & 0.989& 0.761&  0.00 & nan  & nan  \\
0.70 0.30& 0.58 & 0.989& 0.762&  0.42 & 0.989& 0.761  \\
0.70 0.60& 0.18 & 1.000& 0.764&  0.82 & 0.986& 0.761  \\
0.70 0.90& 0.03 & 1.000& 0.785&  0.97 & 0.989& 0.761 \\
 \hline
\end{tabular}
\end{table}

\noindent \textbf{Breast Cancer Diagnosis}:  This task is to map 30 features computed from a digitized image of a fine needle aspirate (FNA) of a breast mass to Class  ``benign'' and  Class ``malignant''~\cite{breast}.  

\begin{table}[h]
\caption{Breast Cancer Diagnosis: NNs}
\label{fig:bcd_nn}
\scriptsize
\center
\begin{tabular}{|l|l|l|l|l|l|l|l|l|}
  \hline
\multicolumn{1}{c}{ $t_1~t_2$}&\multicolumn{3}{c}{$G_1$} &  \multicolumn{3}{c}{$G_2$}\\
  \hline
&  Proportion & Accuracy  &  $\phi_M(x)$   & Proportion&  Accuracy & $\phi_M(x)$ \\
0.80 0.00& 1.00 &  0.892& 0.990& 0.00& nan  & nan  \\
0.80 0.20& 0.75 &  0.916& 0.992& 0.25& 0.821& 0.981  \\
0.80 0.40& 0.48 &  0.907& 0.994& 0.52& 0.879& 0.986  \\
0.80 0.60& 0.26 &  0.926& 0.993& 0.74& 0.881& 0.988  \\
0.90 0.00& 1.00 &  0.899& 0.994& 0.00& nan  & nan  \\
0.90 0.20& 0.76 &  0.914& 0.996& 0.24& 0.849& 0.990  \\
0.90 0.40& 0.48 &  0.905& 0.997& 0.52& 0.892& 0.992  \\
0.90 0.60& 0.26 &  0.924& 0.997& 0.74& 0.890& 0.993  \\
 \hline
\end{tabular}
\end{table}

\begin{table}[h]
\caption{Breast Cancer Diagnosis: SVM}
\label{fig:bcd_svm}
\scriptsize
\center
\begin{tabular}{|l|l|l|l|l|l|l|l|l|}
  \hline
\multicolumn{1}{c}{ $t_1~t_2$}&\multicolumn{3}{c}{$G_1$} &  \multicolumn{3}{c}{$G_2$}\\
  \hline
&  Proportion & Accuracy  &  $\phi_M(x)$   & Proportion&  Accuracy & $\phi_M(x)$ \\
0.80 0.00 &1.00 &  0.988& 0.936 &0.00&nan  & nan  \\
0.80 0.20 &0.77 &  0.995& 0.938 &0.23&0.966& 0.928  \\
0.80 0.40 &0.47 &  0.992& 0.942 &0.53&0.985& 0.930  \\
0.80 0.60 &0.27 &  0.985& 0.944 &0.73&0.989& 0.933  \\
0.90 0.00 &1.00 &  0.995& 0.964 &0.00&nan  & nan  \\
0.90 0.20 &0.78 &  1.000& 0.966 &0.22&0.976& 0.959  \\
0.90 0.40 &0.48 &  1.000& 0.971 &0.52&0.990& 0.958  \\
0.90 0.60 &0.28 &  1.000& 0.969 &0.72&0.993& 0.962  \\
 \hline
\end{tabular}
\end{table}

\begin{table}[h]
\caption{Breast Cancer Diagnosis: GP}
\label{fig:bcd_gp}
\scriptsize
\center
\begin{tabular}{|l|l|l|l|l|l|l|l|l|}
  \hline
\multicolumn{1}{c}{ $t_1~t_2$}&\multicolumn{3}{c}{$G_1$} &  \multicolumn{3}{c}{$G_2$}\\
  \hline
&  Proportion & Accuracy  &  $\phi_M(x)$   & Proportion&  Accuracy & $\phi_M(x)$ \\
0.55 0.00& 1.00 &  0.940& 0.914 &0.00& nan  & nan  \\
0.55 0.20& 0.76 &  0.944& 0.913 &0.24& 0.925& 0.920  \\
0.55 0.40& 0.48 &  0.943& 0.899 &0.52& 0.937& 0.928  \\
0.55 0.60& 0.25 &  0.964& 0.912 &0.75& 0.932& 0.915  \\
0.60 0.00& 1.00 &  0.948& 0.923 &0.00& nan  & nan  \\
0.60 0.20& 0.76 &  0.955& 0.922 &0.24& 0.924& 0.924  \\
0.60 0.40& 0.47 &  0.954& 0.910 &0.53& 0.942& 0.934  \\
0.60 0.60& 0.25 &  0.975& 0.925 &0.75& 0.939& 0.922   \\
 \hline
\end{tabular}
\end{table}

\subsection{Short Summary:}   As we can observe from the experimental results in this section, a strong relationship exists between the ML model's performance and the scores $\phi_{F}(x)$  from FSPT. In regression problems,   we can reduce both  mean error and maximum error, and thereby improve the reliability of ML models, by rejecting predictions with very small  $\phi_{F}(x)$.  In classification problems,  by rejecting input instances with either low  FSPT  score or  ML model  score,  the prediction accuracy can also be improved.  Even among the predictions in which  ML models have similar confidence (i.e., predictions with similar predictive probability $\phi_M(x)$),   the error rate of those with higher $\phi_F(x)$ is  lower than the rest.

\section{Conclusion}

In this paper, we propose  a feature space partition tree (FSPT) to split the feature space into multiple partitions with different training data densities.  The resulting feature space partitions are  scored using a heuristic metric based on the principle that an ML model's performance in a particular feature space partition $\mathcal R$ is upper bounded by the training samples within  $\mathcal R$. \my{ Based on FSPT,  we propose two rejection models for regression and classification problems, respective.} The preliminary experimental results in Section~\ref{sec:evaluation} also meet our expectations.  \my{However, the current version of FSPT has many limitations to be addressed:
\begin{enumerate}
    \item First, the criterion to construct  FSPT  depends on the feature importance or model's reliance on different features. However, it is not a trivial task to get accurate  feature importance values. Besides,  features that are globally important may not be important in the local context, and vice versa. Thus, one possible  direction to improve FSPT is  to incorporate local feature importance in it. 
    \item Another major limitation is that  FSPT is only suitable for low-dimension tabular data sets.  For complex input data such as images,  \my{we should apply FSPT to more meaningful features extracted by other techniques rather than pixel values. For example, DNN trained on images can extract eyes, tail etc. as features in their last layers.}
    \item  \my{Besides,  since the score function is heuristic, we can only show that a strong relationship exists between model performance and FSPT score. In the future, we also plan to derive a more accurate score function.}
    \item  Finally, we also need to  derive a threshold for reject option for a required confidence level. Perhaps, we can apply the conformal prediction framework~\cite{Vovk:2005,vovk2009line} in the different feature space partitions  locally, and derive a threshold for a certain error probability  requirement.
\end{enumerate}

}

\begin{acks}
This work was supported by the Energy Research Institute@NTU.
\end{acks}

\bibliographystyle{ACM-Reference-Format}
\bibliography{main}

\end{document}